\documentclass[aps,prd,onecolumn,groupedaddress,showpacs,nofootinbib,amssymb]{revtex4}
\usepackage{graphicx,bm,color}
\usepackage{amsmath}
\usepackage{amssymb}
\usepackage{amsfonts}

\newcommand{\be}{\begin{equation}}
\newcommand{\ee}{\end{equation}}
\newcommand{\bea}{\begin{eqnarray}}
\newcommand{\eea}{\end{eqnarray}}
\newcommand{\beaa}{\begin{eqnarray*}}
\newcommand{\eeaa}{\end{eqnarray*}}

\newcommand{\nn}{\nonumber \\}
\newcommand{\e}{\mathrm{e}}
\newcommand{\tr}{\mathrm{tr}\,}

\newcommand{\tn}{\tilde{n}}
\newcommand{\tD}{\tilde{D}}

\newcommand{\tx}{\tilde{x}}
\newcommand{\bx}{\mathbf{x}}
\newcommand{\by}{\mathbf{y}}
\newcommand{\mG}{\mathcal{G}}
\newcommand{\mU}{\mathcal{U}}
\newcommand{\mV}{\mathcal{V}}
\newcommand{\mL}{\mathcal{L}}
\newcommand{\mC}{\mathcal{C}}
\newcommand{\mF}{\mathcal{F}}
\newcommand{\mH}{\mathcal{H}}
\newcommand{\mR}{\mathcal{R}}
\newcommand{\mP}{\mathcal{P}}
\newcommand{\pb}[1]{\left\{#1\right\}}

\allowdisplaybreaks[4]

\begin{document}

\title{New proposal for non-linear ghost-free massive $F(R)$ gravity: cosmic
acceleration and Hamiltonian analysis}

\author{Josef Kluso\v{n}$^1$, Shin'ichi Nojiri$^{2,3}$,
and Sergei D. Odintsov$^{4,5,6}$}

\affiliation{
$^1$ Department of Theoretical Physics and Astrophysics,
Faculty of Science, Masaryk University,
Kotl\'{a}\v{r}sk\'{a} 2, 611 37, Brno, Czech Republic \\
$^2$ Department of Physics, Nagoya University, Nagoya
464-8602, Japan \\
$^3$ Kobayashi-Maskawa Institute for the Origin of Particles and
the Universe, Nagoya University, Nagoya 464-8602, Japan \\
$^4$Instituci\`{o} Catalana de Recerca i Estudis Avan\c{c}ats
(ICREA), Barcelona, Spain \\
$^5$Institut de Ciencies de l'Espai
(IEEC-CSIC), Campus
UAB,  Torre C5-Par-2a pl, E-08193 Bellaterra
(Barcelona), Spain \\
$^6$ Tomsk State Pedagogical
University, Tomsk, Russia}

\begin{abstract}
We propose new version of massive $F(R)$ gravity which is natural
generalization of convenient massive ghost-free gravity. Its
Hamiltonian formulation in scalar-tensor frame is developed. We show
that such $F(R)$ theory is ghost-free.
The cosmological evolution of such theory is investigated.
Despite the strong  Bianchi identity constraint the possibility of cosmic
acceleration (especially, in the presence of cold dark matter) is
established.
Ghost-free massive $F(R,T)$ gravity is also proposed.

\end{abstract}


\maketitle

\section{Introduction}

The celebrated theory of the free massive gravity was established about
seventy-five years ago in Ref.~\cite{Fierz:1939ix} (for recent
review, see \cite{Hinterbichler:2011tt}). On the other hand, it has
been known that the interacting or non-linear massive gravity
contains the Boulware-Deser ghost
\cite{Boulware:1974sr,Boulware:1973my} and there appears so-called
vDVZ discontinuity \cite{vanDam:1970vg} in the limit of $m\to 0$.
This discontinuity can be screened by the Vainstein mechanism
\cite{Vainshtein:1972sx} as shown, for example, in
Ref.~\cite{Luty:2003vm} on the example of the DGP model
\cite{Dvali:2000hr}.

Recently, there have been much progress in the non-linear massive gravity study
and the ghost-free construction has been found for non-dynamical background
metric in \cite{deRham:2010ik,deRham:2010kj,Hassan:2011hr} and for dynamical
metric \cite{Hassan:2011zd}.
General proof of absence of ghost in massive gravity has been given in
Ref.~\cite{Hassan:2011tf}.
Especially, the minimal model was first treated in \cite{Hassan:2011vm}.

Cosmological evolution of such massive gravity models has been
investigated in Refs.~\cite{Kluson:2012zz,Kluson:2012wf,Hassan:2011ea}. In case
of bimetric gravity, which contains two metrics
(symmetric tensor fields) some cosmological solutions have been
investigated and the solutions describing accelerating universe are
known
\cite{Damour:2002wu,Volkov:2011an,vonStrauss:2011mq,Berg:2012kn,Nojiri:2012zu,Nojiri:2012re}.
In Ref.~\cite{Nojiri:2012zu,Nojiri:2012re}, we have proposed massive
ghost-free $F(R)$ bigravity model which leads to rich variety of accelerating
universes.  $F(R)$ massive
gravity without dynamical background metric was proposed recently in
\cite{Cai:2013lqa}. In this paper, we propose another kind of
$F(R)$ extension of the massive gravity. We show that due to the
lack of the covariance, the Bianchi identity gives an equation which
constrains the cosmic evolution very strongly. In spite of the
constraint, there appears interesting solution which shows the self-
acceleration when the cold dark matter exists. In this respect, the model is
richer than that of Ref.~\cite{Cai:2013lqa} where spatially-flat FRW expansion
seems to be impossible as we show below.

For consistency of any massive gravity it is crucial to show the
absence of Boulware-Deser ghost \cite{Boulware:1973my}. The general
proof of the absence of the ghost in non-linear massive gravity was
presented in \cite{Hassan:2011ea} and in its St\"{u}ckelberg
formulation in \cite{Hassan:2012qv,Kluson:2012wf}, see also
\cite{Golovnev2:2011aa}. In case of the models studied in this paper,
we also have to show that the Boulware-Deser ghost is absent. We
proceed in the similar way as in \cite{Hassan:2011ea} and we prove
that the Boulware-Deser ghost is absent in all models studied in
this paper\footnote{For similar analysis, see
\cite{Huang:2013mha}.}. More precisely, we use the well-known
equivalence between $f(R)$ theories and scalar-tensor theories to
map the proposed model to the frame where the gravitational action
has the canonical form and where the additional scalar field is
present. It turns out, however, that the presence of this scalar does
not modify the analysis performed in \cite{Hassan:2011ea} and we are
able to show an existence of two  second class constraints that
are crucial for the elimination of the Boulware-Deser ghost and its
conjugate momenta. Then we generalize given analysis to the case of
$f(R,T)$ gravity \cite{Harko:2011kv} with massive term. Introducing
the appropriate number of auxiliary fields we can map given theory
to the non-linear massive gravity coupled to the scalar field and to
the trace of the stress-energy tensor of some matter field. The subsequent
analysis depends on the fact whether we consider $T$ as fixed
parameter or as dynamical quantity. In the second case we have to
introduce the action for the matter as well when we consider
concrete example of the scalar matter field. However, it turns out
that again these theories are ghost free due to the presence of two
second class constraints. This is very remarkable fact since now we
have a possibility to construct broad class of ghost-free non-linear massive
theories of gravity.

\section{Non-linear massive  extension of $F(R)$ gravity}

Let us  propose new model of massive gravity which is an extension of
$F(R)$ gravity (for recent general review, see
\cite{Nojiri:2010wj,Capozziello:2010zz}). The action is given by
\be
\label{mFR1}
S_\mathrm{mg} = M_g^2\int d^4x\sqrt{-\det g}\,F \left(R^{(g)}
+2m^2 \int d^4x\sqrt{-\det g}\sum_{n=0}^{4} \beta_n\,
e_n \left(\sqrt{g^{-1} f} \right) \right)+ S_\mathrm{matter}\, .
\ee
Here $R^{(g)}$ is the scalar curvature for $g_{\mu \nu}$ and
$f_{\mu \nu}$ is a non-dynamical reference metric.
The action for matter is expressed by $S_\mathrm{matter}$.
The tensor $\sqrt{g^{-1} f}$ is defined by the square root of
$g^{\mu\rho} f_{\rho\nu}$, that is,
$\left(\sqrt{g^{-1} f}\right)^\mu_{\ \rho} \left(\sqrt{g^{-1}
f}\right)^\rho_{\ \nu} = g^{\mu\rho} f_{\rho\nu}$.
For general tensor $X^\mu_{\ \nu}$, $e_n(X)$'s are defined by
\begin{align}
\label{ek}
& e_0(X)= 1  \, , \quad
e_1(X)= [X]  \, , \quad
e_2(X)= \tfrac{1}{2}([X]^2-[X^2])\, ,\nn
& e_3(X)= \tfrac{1}{6}([X]^3-3[X][X^2]+2[X^3])
\, ,\nn
& e_4(X) =\tfrac{1}{24}([X]^4-6[X]^2[X^2]+3[X^2]^2
+8[X][X^3]-6[X^4])\, ,\nn
& e_k(X) = 0 ~~\mbox{for}~ k>4 \, .
\end{align}
Here $[X]$ expresses the trace of arbitrary tensor
$X^\mu_{\ \nu}$: $[X]=X^\mu_{\ \mu}$.

In the following, for simplicity, we only consider the minimal case,
\be
\label{mFR2}
2m^2 \sum_{n=0}^{4} \beta_n\,
e_n \left(\sqrt{g^{-1} f} \right)
= 2m^2 \left( 3 - \tr \sqrt{g^{-1} f}+ \det \sqrt{g^{-1} f} \right)\, .
\ee
Then the equation given by the variation of the metric has the following form:
\begin{align}
\label{mFR3}
0 =& M_g^2 \left( \frac{1}{2} g_{\mu\nu} F\left( {\tilde R}^{(g)} 
\right) - R^{(g)}_{\mu\nu} F'\left( {\tilde R}^{(g)} \right)
+ \nabla_\nu \nabla_\mu F'\left( {\tilde R}^{(g)} 
\right) - g_{\mu\nu} \nabla^2 F'\left( {\tilde R}^{(g)} \right) \right) \nn
& + m^2 M_g^2 F'\left( {\tilde R}^{(g)} \right)\left\{
\frac{1}{2} f_{\mu\rho} \left( \sqrt{ g^{-1} f } 
\right)^{-1\, \rho}_{\qquad \nu}
+ \frac{1}{2} f_{\nu\rho} \left( \sqrt{ g^{-1} f } 
\right)^{-1\, \rho}_{\qquad \mu}  - g_{\mu\nu} \det \sqrt{g^{-1} f}
\right\} 
+ \frac{1}{2} T_{\mathrm{matter}\, \mu\nu} \, .
\end{align}
Here  
\be
\label{mFR4}
{\tilde R}^{(g)} \equiv
R^{(g)} + 2m^2 \left( 3 - \tr \sqrt{g^{-1} f}+ \det \sqrt{g^{-1} f} \right)\, ,
\ee
and $\nabla_\mu$ is a covariant derivative given in terms of the Levi-Civita 
connection defined by the metric $g_{\mu\nu}$. In this paper, we do not use the 
covariant derivative with respect to the metric $f_{\mu\nu}$.  
We now have
\begin{align}
\label{mFbis1}
& \nabla^\mu \left( \frac{1}{2} g_{\mu\nu} F\left( {\tilde R}^{(g)} 
\right) - R^{(g)}_{\mu\nu} F'\left( {\tilde R}^{(g)} \right)
+ \nabla_\nu \nabla_\mu F'\left( {\tilde R}^{(g)} 
\right) - g_{\mu\nu} \nabla^2 F'\left( {\tilde R}^{(g)} \right) \right) \nn
=& m^2 F'\left( {\tilde R}^{(g)} \right)
\partial_\nu \left( - \tr \sqrt{g^{-1} f}+ \det \sqrt{g^{-1} f} \right) \, ,
\end{align}
which can be explicitly shown as follows:
\begin{align}
\label{mFbis1b}
& \nabla^\mu \left( \frac{1}{2} g_{\mu\nu} F\left( {\tilde R}^{(g)} 
\right) - R^{(g)}_{\mu\nu} F'\left( {\tilde R}^{(g)} \right)
+ \nabla_\nu \nabla_\mu F'\left( {\tilde R}^{(g)} 
\right) - g_{\mu\nu} \nabla^2 F'\left( {\tilde R}^{(g)} \right) \right) \nn
= & \frac{1}{2} g_{\mu\nu} \left( \nabla^\mu R^{(g)} \right)
F' \left( {\tilde R}^{(g)} \right)
+ m^2 F'\left( {\tilde R}^{(g)} \right)
\partial_\nu \left( - \tr \sqrt{g^{-1} f}+ \det \sqrt{g^{-1} f} 
\right) - \left( \nabla^\mu R^{(g)}_{\mu\nu}\right) 
F'\left( {\tilde R}^{(g)} \right) \nn
& - R^{(g)}_{\mu\nu} \nabla^\mu F'\left( {\tilde R}^{(g)} \right)
+ \nabla^\mu \nabla_\nu \nabla_\mu F'\left( {\tilde R}^{(g)} 
\right) - g_{\mu\nu} \nabla^\mu \nabla^2 F'\left( {\tilde R}^{(g)} \right) \nn
= & \left( \nabla^\mu \left( \frac{1}{2} g_{\mu\nu} R^{(g)} - R^{(g)}_{\mu\nu}
\right) \right) F'\left( {\tilde R}^{(g)} \right)
+ m^2 F'\left( {\tilde R}^{(g)} \right)
\partial_\nu \left( - \tr \sqrt{g^{-1} f}+ \det \sqrt{g^{-1} f} 
\right) - R^{(g)}_{\mu\nu} \nabla^\mu F'\left( R^{(g)} \right) \nn
& - R^{(g) \rho \  \mu}_{\ \ \ \ \mu \  \nu} \nabla_\rho 
F'\left( {\tilde R}^{(g)} \right) + \nabla_\nu \nabla^\mu \nabla_\mu 
F'\left( {\tilde R}^{(g)} \right) - \nabla_{\nu} 
\nabla^2 F'\left( {\tilde R}^{(g)} \right) \nn
= & m^2 F'\left( {\tilde R}^{(g)} \right)
\partial_\nu \left( - \tr \sqrt{g^{-1} f}+ \det \sqrt{g^{-1} f} \right) \, .
\end{align}
Here we have used the Bianchi identity
$0=\nabla^\mu\left( \frac{1}{2} g_{\mu\nu} R^{(g)} - R^{(g)}_{\mu\nu}
\right)$.
Then by multiplying the covariant derivative $\nabla^\mu$ with respect to the
metric $g$ with Eq.~(\ref{mFR3}) and using the conservation law
$0=\nabla^\mu T_{\mathrm{matter}\, \mu\nu} $, we obtain
\begin{align}
\label{mFR5}
0 = & F'\left( {\tilde R}^{(g)} \right)
\partial_\nu \left( - \tr \sqrt{g^{-1} f}+ \det \sqrt{g^{-1} f} \right) \nn
&+ \left( \partial^\mu F'\left( {\tilde R}^{(g)} \right) \right)\left\{
\frac{1}{2} f_{\mu\rho} \left( \sqrt{ g^{-1} f } \right)^{-1\, \rho}_{\qquad
\nu}
+ \frac{1}{2} f_{\nu\rho} \left( \sqrt{ g^{-1} f } \right)^{-1\, \rho}_{\qquad
\mu}
   - g_{\mu\nu} \det \sqrt{g^{-1} f}
\right\} \nn
& + F'\left( {\tilde R}^{(g)} \right) \nabla^\mu \left\{
\frac{1}{2} f_{\mu\rho} \left( \sqrt{ g^{-1} f } \right)^{-1\, \rho}_{\qquad
\nu}
+ \frac{1}{2} f_{\nu\rho} \left( \sqrt{ g^{-1} f } \right)^{-1\, \rho}_{\qquad
\mu}
   - g_{\mu\nu} \det \sqrt{g^{-1} f}
\right\} \, .
\end{align}
We now assume the FRW universe for the metrics $g_{\mu\nu}$ and flat Minkowski
space-time
for $f_{\mu\nu}$ and use the conformal time $t$ for the universe with metric
$g_{\mu\nu}$:
\be
\label{Fbi10}
ds_g^2 = \sum_{\mu,\nu=0}^3 g_{\mu\nu} dx^\mu dx^\nu
= a(t)^2 \left( - dt^2 + \sum_{i=1}^3 \left( dx^i \right)^2\right) \, ,\quad
ds_f^2 = \sum_{\mu,\nu=0}^3 f_{\mu\nu} dx^\mu dx^\nu
= - dt^2 + \sum_{i=1}^3 \left( dx^i \right)^2 \, .
\ee
Then the $\nu=i$ component in (\ref{mFR5}) is trivially satisfied.
On the other hand, the $\nu=t$ component gives
\be
\label{mFR6}
0 = \partial_t \left( - 4 a^{-1} + a^{-4} \right) F'\left( {\tilde R}^{(g)}
\right)
+ \left( a^{-1} - a^{-4} \right) \partial_t F'\left( {\tilde R}^{(g)} \right)
= a^{-4} \partial \left\{ F' \left({\tilde R}^{(g)} \right) \left( a^3 - 1
\right) \right\}\, ,
\ee
which further gives
\be
\label{mFR7}
F'\left( {\tilde R}^{(g)} \right) \left( a^3 - 1 \right) = C\, ,
\ee
with a constant $C$.
Eq.~(\ref{mFR7}) determines the form of $F'\left( {\tilde R}^{(g)} \right)$.
For a given time evolution of the scale factor $a=a(t)$, we find the $t$
dependence of ${\tilde R}^{(g)}$:
${\tilde R}^{(g)}={\tilde R}^{(g)}(t)$, which can be solved with respect to $t$
as a function of
${\tilde R}^{(g)}$: $t = t \left( {\tilde R}^{(g)} \right)$. Then
Eq.~(\ref{mFR7}) gives the form of
$F'\left( {\tilde R}^{(g)} \right)$ as follows,
\be
\label{mFR8}
F'\left( {\tilde R}^{(g)} \right)
= \frac{C}{ a\left(t \left( {\tilde R}^{(g)} \right)\right)^3 - 1 } \, ,
\ee
As we will see soon, however, the time evolution of the scale factor
$a=a(t)$ cannot be arbitrary.
We should also note that $F'\left( {\tilde R}^{(g)} \right)$ diverges when the
scale factor $a$
goes to unity.

In the FRW metric with conformal time in (\ref{Fbi10}),
the $(\mu,\nu)=(t,t)$ component in (\ref{mFR3}) has the following form:
\be
\label{mFR9}
0 = - \frac{1}{2} a^{-2} F\left( {\tilde R}^{(g)} \right)
+ 3 \dot H F'\left( {\tilde R}^{(g)} \right)
   - 3 H \partial_t F'\left( {\tilde R}^{(g)} \right)
+ \left( - a + a^{-3} \right) F'\left( {\tilde R}^{(g)} \right)
+ \frac{1}{2 M_g^2}\rho_\mathrm{matter}\, ,
\ee
and the $(\mu,\nu)=(i,j)$ component gives
\be
\label{mFR10}
0 = \frac{1}{2} a^{-2} F\left( {\tilde R}^{(g)} \right)
   -\left( \dot H + 2 H^2 \right) F'\left( {\tilde R}^{(g)} \right)
+ \left( \partial_t^2 + H \partial_t \right) F'\left( {\tilde R}^{(g)} \right)
   - \left( - a + a^{-3} \right) F'\left( {\tilde R}^{(g)} \right)
+ \frac{1}{2 M_g^2}p_\mathrm{matter}\, .
\ee
By combining (\ref{mFR9}) and (\ref{mFR10}), one obtains
\be
\label{mFR11}
0 = 2 \left( \dot H - H^2 \right) F'\left( {\tilde R}^{(g)} \right)
+ \left( \partial_t^2 - 2 H \partial_t \right) F'\left( {\tilde R}^{(g)}
\right)
+ \frac{1}{2 M_g^2}\left( \rho_\mathrm{matter} + p_\mathrm{matter} \right)\, .
\ee
Different from the Einstein gravity, Eqs.~(\ref{mFR9}), (\ref{mFR10}), and
conservation law
\be
\label{mFR12}
0 = \dot \rho_\mathrm{matter} + 3 H \left( \rho_\mathrm{matter} +
p_\mathrm{matter} \right)\, ,
\ee
are independent equations.
The form of the conservation law in terms of the conformal time is not changed
from that of
the cosmological time.
Instead of Eqs.~(\ref{mFR9}), (\ref{mFR10}), and (\ref{mFR12}), we may regard
Eqs.~(\ref{mFR7}),
(\ref{mFR11}), and the conservation law (\ref{mFR12}) as independent equations.
Eq.~(\ref{mFR7}) gives
\begin{align}
\label{mFR13}
\partial_t F'\left( {\tilde R}^{(g)} \right) = - \frac{3H a^3 C}{a^3 - 1} \,
,\quad
\partial_t F'\left( {\tilde R}^{(g)} \right) = \left\{ \frac{18 H^2 a^6}{\left(
a^3 - 1 \right)^3}
   - \frac{\left( 3 \dot H + 9 H^2 \right)a^3}{ \left( a^3 - \right)^2} \right\}
C\, .
\end{align}
Then Eq.~(\ref{mFR11}) can be rewritten as
\be
\label{mFR14}
0 = \frac{\left\{ \dot H \left( - a^6 - a^3 + 2 \right)
+ H^2 \left( 13 a^6 + 7 a^3 - 2 \right) \right\}C}{\left( a^3 - 1 \right)^3}
+ \frac{1}{2 M_g^2}\left( \rho_\mathrm{matter} + p_\mathrm{matter} \right)\, .
\ee
Independent from the form of $F\left( {\tilde R}^{(g)} \right)$,
Eq.~(\ref{mFR14}) describes the dynamics of the
universe.

It is difficult to solve (\ref{mFR14}) explicitly.
Then it is easier to consider the following three cases: a) $a\to 1$ case, b)
$a\gg 1$ case, c) $a\ll 1$ case.
In the following, for simplicity, we assume that the matter has a constant
equation of state (EoS)
parameter $w$ and therefore
\be
\label{mFR14b}
p_\mathrm{matter} = w \rho_\mathrm{matter}\, ,\quad
\rho_\mathrm{matter} = \rho_0 a^{-3 \left(w + 1 \right)}\, .
\ee

\noindent
a) $a\to 1$ case. By putting $a= 1 + \delta a$, from (\ref{mFR14}), we obtain
\be
\label{mFR15}
0 \sim - 9 \dot H + 18 H^2 \sim - 9 \delta\ddot a \delta a + 18 \left( \delta
\dot a \right)^2
= - 9 \left( \delta a \right)^3 \frac{d}{dt} \left( \frac{\delta \dot a}{\left(
\delta a \right)^2 }
\right)\, ,
\ee
whose solution is given by
\be
\label{mFR16}
\delta a = \frac{C_1}{t + C_2}\, .
\ee
Here $C_1$ and $C_2$ are constants of the integration.
Eq.~(\ref{mFR16}) tells us that the limit $a\to 1$ $\left( \delta a \to 0
\right)$ is realized
in the infinite past or future in conformal time, $t \to \pm \infty$.

\noindent
b) $a\gg 1$ case. In this case, Eq.~(\ref{mFR14}) can be approximated as
\be
\label{mFR17}
0 \sim C a^{-3} \left( - \dot H + 13 H^2 \right)
+ \frac{1+w}{2} \rho_0 a^{- 3 \left( w + 1 \right)}\, .
\ee
\begin{enumerate}
\item $\rho_0 = 0$ case. The solution of (\ref{mFR17}) is given by
\be
\label{mFR18}
H = \frac{1}{13 \left( t_0 - t \right)}\, ,
\ee
which describes the phantom universe which has a Big Rip singularity at $t=t_0$
since we assume $a\gg 1$.
\item $\rho_0 \neq 0$ case. If $w\neq 0$, we have a power law solution,
\be
\label{mFR19}
a = a_0 t^{\frac{2}{3w}}\, .
\ee
Here $a_0$ is given by solving the following equation
\be
\label{mFR20}
0 = \frac{2C}{3w} \left( 1 + \frac{26}{3w} \right)
+ \frac{w + 1}{2} a_0^{-3w} \rho_0\, .
\ee
On the other hand, when $w=0$, we obtain a solution describing de Sitter
universe:
\be
\label{FR20b}
H^2 = \frac{\rho_0}{26C}  \, .
\ee
This could be interesting since the accelerating expansion of the present
universe can be realized by dust,
which may be identified with cold dark matter.
Then we find
\be
\label{FR20c}
\frac{1}{26C} \rho_0 \sim \left( 10^{-33}\, \mathrm{eV} \right)^2\, ,\quad
\rho_0 a^3 \sim \left( 10^{-3}\, \mathrm{eV} \right)^4\, .
\ee

\end{enumerate}

\noindent
c) $a\ll 1$ case. The approximated form of Eq.~(\ref{mFR14}) in this case is
given by
\be
\label{mFR21}
0 = - 2 C \left( \dot H - H^2 \right) + \frac{1+w}{2} \rho_0 a^{- 3 \left( w +
1 \right)}\, .
\ee
\begin{enumerate}
\item $\rho_0 = 0$ case. The solution is given by
\be
\label{mFR22}
H = \frac{1}{t_0 - t}\, .
\ee
with a constant of integration $t_0$. The expression (\ref{mFR22}) is valid
when $t \to \pm \infty$
because we are assuming $a\ll 1$. Therefore, in spite of the form in
(\ref{mFR22}), there does not always occur a Big Rip singularity.
\item $\rho_0 \neq 0$ case. By solving (\ref{mFR22}), we find
\be
\label{mFR23}
a = a_0 t^{\frac{2}{3\left( w + 1 \right)}}\, .
\ee
Now $a_0$ is given by solving the following equation:
\be
\label{mFR24}
0 = - \frac{4C}{{3\left( w + 1 \right)}} \left( 1 - \frac{2}{3\left( w + 1
\right)} \right)
+ \frac{w + 1}{2} a_0^{-3w} \rho_0\, .
\ee
The qualitative behavior is not changed from the Einstein gravity coupled with
matter
of a constant EoS parameter $w$.
\end{enumerate}

Thus, we demonstrated the principal possibility to have accelerating cosmology
within new non-linear massive $F(R)$ gravity. Nevertheless,
the variety of possible cosmological solutions is not so wide as in convenient
$F(R)$ gravity.

\section{FRW cosmology from another massive $F(R)$ gravity model}

Instead of the model in (\ref{mFR1}) for the purpose of comparison, we now
consider an $F(R)$ extension
of massive gravity proposed recently in \cite{Cai:2013lqa}.
The action is given by
\be
\label{massivegravity}
S_\mathrm{mg} = M_g^2\int d^4x\sqrt{-\det g}\,F \left(R^{(g)}\right)
+2m^2 M_g^2 \int d^4x\sqrt{-\det g}\sum_{n=0}^{4} \beta_n\,
e_n \left(\sqrt{g^{-1} f} \right) + S_\mathrm{matter}\, .
\ee
In the following, just for simplicity, we only consider the minimal case,
as in (\ref{mFR2}),
\be
\label{mg3}
S_\mathrm{mg} = M_g^2\int d^4x\sqrt{-\det g}\,F \left( R^{(g)} \right)
+2m^2 M_g^2 \int d^4x\sqrt{-\det g} \left( 3 - \tr \sqrt{g^{-1} f}
+ \det \sqrt{g^{-1} f} \right) + S_\mathrm{matter}\, ,
\ee
Then by the variation over $g_{\mu\nu}$, we obtain
\begin{align}
\label{Fbi8}
0 =& M_g^2 \left( \frac{1}{2} g_{\mu\nu} F\left( R^{(g)} \right)
   - R^{(g)}_{\mu\nu} F'\left( R^{(g)} \right)
+ \nabla_\nu \nabla_\mu F'\left( R^{(g)} \right)
   - g_{\mu\nu} \nabla^2 F'\left( R^{(g)} \right) \right) \nn
& + m^2 M_g^2 \left\{ g_{\mu\nu} \left( 3 - \tr \sqrt{g^{-1} f} \right)
+ \frac{1}{2} f_{\mu\rho} \left( \sqrt{ g^{-1} f } \right)^{-1\, \rho}_{\qquad
\nu}
+ \frac{1}{2} f_{\nu\rho} \left( \sqrt{ g^{-1} f } \right)^{-1\, \rho}_{\qquad
\mu}
\right\} 
+ \frac{1}{2} T_{\mathrm{matter}\, \mu\nu} \, .
\end{align}
Now, instead of (\ref{mFbis1}), we have
\be
\label{Fbis1}
0 = \nabla^\mu \left( \frac{1}{2} g_{\mu\nu} F\left( R^{(g)} \right)
   - R^{(g)}_{\mu\nu} F'\left( R^{(g)} \right)
+ \nabla_\nu \nabla_\mu F'\left( R^{(g)} \right)
   - g_{\mu\nu} \nabla^2 F'\left( R^{(g)} \right) \right)\, ,
\ee
and \be
\label{identity1}
0 = - g_{\mu\nu} \nabla^\mu \left( \tr \sqrt{g^{-1} f} \right)
+ \frac{1}{2} \nabla^\mu \left\{ f_{\mu\rho} \left( \sqrt{ g^{-1} f }
\right)^{-1\, \rho}_{\qquad \nu}
+ f_{\nu\rho} \left( \sqrt{ g^{-1} f } \right)^{-1\, \rho}_{\qquad \mu}
\right\} \, ,
\ee
which corresponds to (\ref{mFR5}).
Starting again from the FRW universe for the metric $g_{\mu\nu}$ and flat
Minkowski space-time
for $f_{\mu\nu}$ one can  use the conformal time $t$ as in (\ref{Fbi10}).
Then $(t,t)$ component of (\ref{Fbi8}) gives
\be
\label{Fbi11}
0 = - 3 M_g^2 H^2 - 3 m^2 M_g^2
\left( a^2 - a \right) + \rho_\mathrm{matter} \, ,
\ee
and $(i,j)$ components give
\be
\label{Fbi12}
0 = M_g^2 \left( 2 \dot H + H^2 \right)
+  3 m^2 M_g^2 \left( a^2 - a \right) + p_\mathrm{matter}\, .
\ee
Here $H=\dot a / a$.
Eq.~(\ref{identity1}) gives the following constraint:
\be
\label{identity3}
\frac{\dot a}{a} = 0\, .
\ee
Different from Eq.~(\ref{mFR7}), the identity (\ref{identity3}) shows that
$a$ should be a constant $a=a_0$.
This indicates that the only consistent solution for $g_{\mu\nu}$
is the flat Minkowski space.
Therefore we cannot obtain the expanding universe without extra fields and/or
fluids.

In \cite{Cai:2013lqa}, this model was studied  in the FRW universe with
non-vanishing spatial curvature.
When the spatial curvature does not vanish, the scale factor is proportional to
the spatial curvature
and linear to the cosmological time (not conformal time as in this paper).
In the limit that the spatial curvature vanishes, the scale factor becomes a
constant,
what is consistent with the result obtained here.
Even in case that the spatial curvature does not vanish, Eq.~(\ref{identity1})
gives a strong constraint with  the only possible solution, for which the scale
factor must be linear in cosmological time.


\section{Hamiltonian Formalism}

In this section, we perform Hamiltonian formulation of the model described by
the action (\ref{mFR1}) and show that the model does not contain ghost.
As the first step we introduce two auxiliary fields $A$, $B$ and rewrite the
action
(\ref{mFR1}) into the following form
\begin{equation}\label{actH}
S=M_g^2\int d^4x\sqrt{-\det g}\left[B \left(R^{(g)} + 2m^2 \sum_{n=0}^4 \beta_n
e_n
(\sqrt{- g^{-1} f})-A \right) + F(A)\right] \, .
\end{equation}
Using the Weyl transformation
\begin{equation}\label{Weyltr}
g'_{\mu\nu}=\Omega g_{\mu\nu}
\end{equation}
   implies
\begin{equation}
R^{(g)}\left[g \right] = \Omega \left( R^{(g)}\left[g'
\right]-\frac{3}{2\Omega^2}
g'^{\mu\nu}\nabla'_\mu\Omega
\nabla'_\nu\Omega+3 g'^{\mu\nu} \left(\frac{1}{\Omega}
\nabla'_\mu\nabla'_\nu\Omega-\frac{1}{\Omega^2}\nabla'_\mu \Omega
\nabla'_\nu\Omega\right) \right)
\end{equation}
Here $\nabla'_\mu$ is the covariant derivative with respect to $g'_{\mu\nu}$.
Now by choosing $B=\Omega$, we find the theory with the canonical
Einstein-Hilbert term. Further, the  equation of motion with
respect to $A$ has the form
\begin{equation}
F_A(A)=B \, .
\end{equation}
Here $F_A(A) \equiv d F(A)/ dA$.
We presume that it can be solved for $A=\Psi(B)$, where
$F_A(\Psi(x))=x$.
Then by introducing the scalar field $\phi$ through the formula
$\Omega=\exp (\phi)$, we obtain the action in the form
\be
\label{act2}
S=M_g^2\int d^4x \sqrt{-\det g'}\left[ R^{(g)} -\frac{2}{3}
g'^{\mu\nu}\nabla'_\mu\phi
\nabla'_\nu\phi-V(\phi) \right] + 2m^2 \int d^4x \sqrt{- g'}\sum_{n=0}^4
\beta_n
\e^{\frac{n}{2}\phi} e_n \left( \sqrt{{g'}^{-1}f} \right) \, .
\ee
In what follows we omit $'$ over metric variables.

Our goal is to find the Hamiltonian formulation of given theory and
determine corresponding primary and the secondary constraints. As
the first step we introduce $3+1$ decomposition of
both $g_{\mu\nu}$ and $f_{\mu\nu}$ \cite{Gourgoulhon:2007ue,Arnowitt:1962hi}
\be
g_{00} = -N^2+N_i g^{ij}N_j \, , \quad g_{0i}=N_i \, ,
\quad \hat{g}_{ij}=g_{ij} \, ,\quad
g^{00} = -\frac{1}{N^2} \, , \quad g^{0i}=\frac{N^i}{N^2} \, ,
\quad g^{ij}=g^{ij}-\frac{N^i N^j}{N^2} \, ,
\ee
and
\begin{align}
& f_{00} = -M^2+L_i f^{ij}L_j \, , \quad f_{0i}=L_i \, , \quad
f_{ij}=f_{ij} \, , \nonumber \\
& f^{00} = -\frac{1}{M^2} \, , \quad f^{0i}=-\frac{L^i}{M^2} \, ,
\quad f^{ij}= f^{ij}-\frac{L^i L^j}{M^2} \, , \quad  L^i=L_jf^{ji} \, ,
\end{align}
where we defined $g^{ij}$ and  $f^{ij}$ as the inverse to
$g_{ij}$ and $f_{ij}$, respectively
$g_{ik}g^{kj}=\delta_i^{ \ j}$, $f_{ik}f^{kj}=\delta_i^{ \ j}$.
By following \cite{Hassan:2011zd,Hassan:2011tf,Hassan:2011hr,Hassan:2011vm},
we perform the following redefinition of the shift function
\begin{equation}
N^i=M\tn^i+L^i+N \tD^i_{ \ j } \tn^j \, ,
\end{equation}
so that the resulting action is linear in $M$ and $N$.
Note that the matrix $\tD^i_{ \ j}$ obeys the equation
\cite{Hassan:2011zd,Hassan:2011tf,Hassan:2011hr,Hassan:2011vm}
\begin{equation}
\label{defD}
\sqrt{\tx}
\tD^i_{ \ j}= \sqrt{( g^{ik}-\tD^i_{ \ m} \tn^m \tD^k_{ \ n}\tn^n )f_{kj}}
\end{equation}
and also following important property
$f_{ik}\tD^k_{ \ j} = f_{jk}\tD^k_{ \ i }$.
Then after some calculations, we find that the action (\ref{act2})
has the form
\begin{align}
\label{SRbi}
S =& M_g^2\int dt d^3\bx N\sqrt{^{(3)}g}[K_{ij}\mG^{ijkl}K_{kl}+{}^{(3)}R^{(g)}
+\frac{2}{3}\nabla_n\phi\nabla_n\phi-\frac{2}{3}g^{ij}
\partial_i\phi\partial_j\phi-V(\phi)
] \nonumber \\
&+ 2m^2M_\mathrm{eff}^2 \int  dt d^3\bx \sqrt{^{(3)}g} (M \mU+N\mV) \, ,
\end{align}
where
\be
K_{ij} = \frac{1}{2N}(\partial_t g_{ij}- \nabla_i
N_j(\tn,g)-\nabla_j N_i(\tn,g)) \, , \quad
\nabla_n\phi = \frac{1}{N}(\partial_t\phi-N^i(\tn,g)\partial_i\phi) \, ,
\ee
with
\begin{equation}
N_i=M g_{ij}\tn^j+g_{ij}L^j+Ng_{ik}\tD^k_{ \ j}\tn^j \, , \quad
L_i=f_{ij}L^j \, ,
\end{equation}
and  where $\nabla_i,{}^{(3)} R^{(g)}$  are the covariant derivative and
scalar curvature  calculated using $g_{ij}$.
We should also note that $^{(3)}g$ is the determinant of $g_{ij}$.
Furthermore, $\mG^{ijkl}$ are the de Witt metrics defined as
\begin{equation}
\mG^{ijkl}=\frac{1}{2}(g^{ik}g^{jl}+g^{il}g^{jk})-g^{ij}g^{kl} \, ,
\quad
\end{equation}
with inverse
\begin{equation}
\mG_{ijkl}=\frac{1}{2}(g_{ik}g_{jl}+ g_{il}g_{jk})-\frac{1}{2}
g_{ij}g_{kl} \, ,
\end{equation}
that obey the relation
\begin{equation}
\mG_{ijkl}\mG^{klmn}=\frac{1}{2}(\delta_i^m\delta_j^n+
\delta_i^n\delta_j^m)    \, .
\end{equation}
Finally, $ \mV$ and $\mU$ introduced  in (\ref{SRbi}) have the following form
\begin{align}
\mV =& \beta_0 +\beta_1\e^{\frac{1}{2}\phi} \sqrt{\tx}\tD^i_{ \ i}
+\beta_2\e^{\phi}\frac{1}{2}\sqrt{\tx}^2 \left(\tD^i_{ \ i}\tD^j_{ \ j} -
\tD^i_{ \ j}
\tD^j_{ \ i} \right)
+\frac{1}{6}\beta_3 \e^{\frac{3}{2}\phi}  \sqrt{\tx}^3 \left[ \tD^i_{ \ i}
\tD^j_{ \ j}\tD^k_{ \ k}-3\tD^i_{ \ i}\tD^j_{ \ k} \tD^k_{ \ j}
+2\tD^i_{ \ j}\tD^j_{ \ k}\tD^k_{ \ i} \right] \, ,
\nonumber \\
\mU =& \beta_1 \e^{\frac{1}{2}\phi} \sqrt{\tx}+ \beta_2  \e^{\phi}
\left[ \sqrt{\tx}^2 \tD^i_{ \ i}+\tn^i f_{ij}\tD^j_{ \ k}\tn^k \right]
\nonumber \\
& + \beta_3 \e^{\frac{3}{2}\phi} \left[ \sqrt{\tx}(\tD^l_{ \ l}\tn^i f_{ij}
\tD^j_{ \ k}\tn^k- \tD^i_{ \ k}\tn^k f_{ij}\tD^j_{ \ l}\tn^l)
+ \frac{1}{2} \sqrt{\tx}^3 (\tD^i_{ \ i}\tD^j_{ \ j}-\tD^i_{ \ j}\tD^j_{ \ i})
\right]
+ \beta_4 \e^{2\phi} \frac{\sqrt{^{(3)}f}}{\sqrt{^{(3)}g}} \, ,
\end{align}
where $\tx=1-\tn^if_{ij}\tn^j$ and $^{(3)}f$ is the determinant of $f_{ij}$.
The action (\ref{SRbi}) is suitable for the Hamiltonian formalism. First we
find the momenta conjugate to $N,\tn^i$ and $g_{ij}$
\begin{equation}
\pi_N\approx 0 \, , \quad  \pi_i\approx 0 \, , \quad
\pi^{ij}=M_g^2\sqrt{^{(3)}g}\mG^{ijkl}K_{kl} \, ,
\end{equation}
together with the momenta conjugate to $\phi$
\begin{equation}
p_\phi=\frac{4}{3}M_g^2\sqrt{^{(3)}g}\nabla_n\phi \, .
\end{equation}
Then after some calculations we find the following Hamiltonian
\be
H = \int d^3\bx \left( \pi^{ij}\partial_t g_{ij}+ p_\phi
\partial_t \phi-\mL \right)
= \int d^3\bx \left( N\mC_0+\mH_0 \right)\, ,
\ee
where
\begin{align}
\mC_0 =& \frac{1}{M_g^2\sqrt{^{(3)}g}} \pi^{ij}
\mG_{ijkl}\pi^{kl}-M_g^2\sqrt{^{(3)}g}{}^{(3)}
R+(\mR_k+p_\phi\partial_k\phi)\tD^k_{ \ l}\tn^l -2m^2\sqrt{^{(3)}g}\mV
\nonumber \\
& +
\frac{3}{8}\frac{1}{\sqrt{^{(3)}g}M_g^2}p_\phi^2+\frac{2}{3}M_g^2\sqrt{^{(3)}g}g^{ij}
\partial_i\phi\partial_j\phi+M_g^2\sqrt{^{(3)}g}V(\phi) \, ,  \nonumber \\
\mH_0=& (M\tn^i+L^i)(\mR_i+p_\phi\partial_i\phi)-2m^2M\sqrt{^{(3)}g} \mU
\, ,
\end{align}
where we also denoted $\mR_i=-2g_{ik}\nabla_l\pi^{lk}$.
We see that the theory possesses four primary constraints
\begin{equation}
\pi_N\approx 0 \, , \quad \pi_i\approx 0 \, ,
\end{equation}
where $\pi_N$ and $\pi_i$ are momenta conjugate to $N$ and $\tn^i$,
respectively with the following non-zero Poisson brackets
\begin{equation}
\pb{N(\bx),\pi_N(\by)}=\delta(\bx-\by) \, , \quad
\pb{\tn^i(\bx),\pi_j(\by)}=\delta^i_j\delta(\bx-\by) \, .
\end{equation}
To proceed further we need following relations
\begin{align}
\frac{\delta \sqrt{\tx}\tD^k_{ \ k}}{\delta \tn^i}
=& -\frac{1}{\sqrt{\tx}}\tn^pf_{pk}\frac{\delta }{\delta \tn^i}
(\tD^k_{ \ m}\tn^m) \, .\nonumber \\
\frac{\delta}{\delta \tn^i}\tr \left( \sqrt{\tx}\tD\sqrt{\tx}\tD \right)
=& -2 \tn^jf_{jk}\tD^k_{ \ l}\frac{\delta \left( \tD^l_{ \ m}\tn^m
\right)}{\delta
\tn^i} \, , \nonumber \\
\frac{\delta}{\delta \tn^i} \tr
\left(\sqrt{\tx}\tD\sqrt{\tx}\tD\sqrt{\tx}\tD\right)
=& -3\sqrt{\tx} \tn^j
f_{jk}\tD^j_{ \ m}\tD^m_{ \ n}\frac{\delta \left(\tD^n_{ \ p}\tn^p \right)}{
\delta \tn^i} \, ,
\end{align}
that follow from (\ref{defD}) and also using the property
$f_{ik}\tD^k_{ \ j}=f_{jk}\tD^k_{ \ i}$.
Then  we find
\begin{align}
\frac{\delta \mH_0}{\delta \tn^i}=& M \left(\mR_i+p_\phi
\partial_i\phi \right) \nn
& +2m^2M\sqrt{^{(3)}g} \left[ \frac{\beta_1}{\sqrt{\tx}}
\e^{\frac{1}{2}\phi}f_{ij}\tn^j+\beta_2 \e^\phi \left(f_{ij}\tn^j
\tD^i_{ \ i} - f_{ij}\tD^i_{ \ k}\tn^k \right) \right. \nn
& \left. +\beta_3 n^pf_{pj} \e^{\frac{3}{2}}\sqrt{\tx} \left(
\frac{1}{2}\delta^j_i(\tD^m_{ \ m} \tD^n_{ \ n}-\tD^m_{ \ n}\tD^n_{\ m} \right)
+ \tD^j_{ \ m}\tD^m_{ \ i} -\tD^j_{ \ i}\tD^m_{ \ m}) \right] \nn
\equiv& M\mC_i \, ,
\end{align}
where $\mC_i$ is defined by
\begin{align}
\mC_i = & \mR_i+p_\phi\partial_i\phi
+ 2m^2\sqrt{^{(3)}g}\frac{f_{ij}\tn^j}{\sqrt{\tx}}\left[\beta_1
\e^{\frac{1}{2}\phi}\delta^j_{ \ i}+\beta_2 \e^\phi
\sqrt{\tx} \left(\delta^j_{ \ i}\tD^m_{ \ m}-\tD^j_{ \ i}\right) \right.
\nonumber \\
& +  \left. \beta_3 \left(\sqrt{\tx}\right)^2
\e^{\frac{3}{2}}
\left(
\frac{1}{2}\delta^j_i \left(\tD^m_{ \ m} \tD^n_{ \ n}-\tD^m_{ \ n}\tD^n_{ \
m}\right)
+ \tD^j_{ \ m}\tD^m_{ \ i} -\tD^j_{ \ i}\tD^m_{ \ m}\right)
\right] \, .
\end{align}
In the same way, we find
\be
\frac{\delta \mC_0}{\delta \tn^i}=\mC_j
\frac{\delta \left( \tD^j_{ \ k}\tn^k \right)}{\delta \tn^i} \, .
\ee
Now the requirement of the preservation of the primary constraints
$\pi_N\approx 0$, $\pi_i\approx 0$ implies
\begin{align}
\partial_t\pi_N =&\pb{\pi_N,H}=-\mC_0\approx 0 \, ,
\nonumber \\
\partial_t\pi_i =&\pb{\pi_i,H}=-\frac{\delta H}{\delta \tn^i}=
-\mC_j\left(M\delta^j_i+\frac{\delta (\tD^j_{ \ k}\tn^k)}{\delta
\tn^i}\right) =0 \, ,
\end{align}
that implies an existence of the secondary constraints $\mC_i\approx
0$. As a result we find that the total Hamiltonian has the following form
\begin{equation}
H_T=\int d^3\bx \left(\mH_0+N\mC_0+v_N\pi_N+v^i\pi_i+\Sigma^i\mC_i \right) \,
\end{equation}
where $v_N,v^i,\Sigma^i$ are Lagrange multipliers corresponding to the
constraints
$\pi_N\approx 0, \pi_i\approx 0, \mC_i\approx 0$.

The next step is to analyze the requirement of the preservation of
the secondary constraints $\mC_0$, $\mC_i$. First of all note that the
requirement of the preservation of the constraint $\pi_i\approx 0$
implies
\begin{equation}
\partial_t\pi_i=\pb{\pi_i,H_T}\approx
\int d^3\bx \Sigma^j\pb{\pi_i,\mC_j}=0 \, .
\end{equation}
It can be shown that $\pb{\pi_i(\bx),\mC_j(\by)}\equiv
\triangle_{ij}\delta(\bx-\by)$ where $\triangle_{ij}$ is
a non-singular matrix so that the only possible solution of the
equation above is $\Sigma^j=0$.

Now we have to determine the time evolution of the constraint
$\mC_0$:
\be
\partial_t\mC_0 = \pb{\mC_0,H_T}
=\int d^3\bx \left[ N(\bx)\pb{\mC_0,\mC_0(\bx)} +\pb{\mC_0,\mH(\bx)} \right] =0
\, .
\ee
To proceed further  we determine the Poisson bracket
$\pb{\mC_0(\bx),\mC_0(\by)}$. Following \cite{Hassan:2011ea}, we
easily find that this Poisson bracket has the following form:
\begin{equation}
\pb{\mC_0(\bx),\mC_0(\by)}=-\mP^i(\by)\frac{\partial}{\partial y^i}
\delta(\bx-\by)+\mP^i(\bx)\frac{\partial}{\partial
x^i}\delta(\bx-\by) \, ,
\end{equation}
where
\begin{equation}
\mP^i=\mC_0\tD^i_{ \ k}\tn^k+\mC_jg^{ji} \, .
\end{equation}
Since $\mP^i$ is given as the linear combination of the constraints,
we find that $\mP^i$ vanishes on the constraint surface so that
\begin{equation}
\pb{\mC_0(\bx),\mC_0(\by)}\approx 0 \, .
\end{equation}
Now it is easy to see that the requirement of the preservation of
the constraint $\mC_0$ implies following secondary constraint
\begin{equation}
\int d^3\bx \pb{\mC_0,\mH(\bx)}\equiv \mC^{(\mathrm{II})} \approx 0 \, ,
\end{equation}
with explicit form that it is not important for us.
Finally the
requirement of the preservation of the constraint $\mC_i$ takes the
form
\be
\partial_t\mC_i=\pb{\mC_i,H_T}=
\int d^3\bx \left(
\pb{\mC_i,\mH(\bx)}+ v^j\pb{\mC_i,\pi_j(\by)} \right)\approx 0\, ,
\nonumber \\
\ee that using the same arguments as in case of the preservation of
the constraint $\pi_i\approx 0$ implies that given equation can be
solved for $v^j$ as functions of canonical variables. Note also that
$\pi_N\approx 0$ is the first class constraint. Finally it can be
easily shown that $\mC,\mC^{(\mathrm{II})}$ has non-trivial Poisson
bracket which implies that the are the second class constraints
\cite{Hassan:2011ea}.

In
summary we have following structure of constraints. We have six
second class constraints $\pi_i\approx 0$, $\mC_i\approx 0$ that can be
solved for $\pi_i$ and for $\tn^i$. Then we have one first class
constraint $\pi_N$ that can be gauge fixed by imposing the condition
$N=1$ (for example). Finally $\mC$, $\mC^{(\mathrm{II})}$ are the second class
constraints that can be solved for the Boulware-Deser ghost and its
conjugate momenta. As a result we find that this theory possesses
$12$ degrees of freedom where $10$ of them correspond to the massive
gravity and $2$ corresponds to $\phi$, $p_\phi$.

We can generalize the ghost-free proposal  (\ref{actH}) in several ways. For
example, let us
consider non-linear massive theory where $F(R)$ depends on the trace
of the stress energy tensor \cite{Harko:2011kv}
\be
\label{SactA1}
S=M_g^2\int d^4x \sqrt{-\det g}F \left( R^{(g)},T \right) +2m^2 \int
d^4x\sqrt{-\det g}\sum_{n=0}^4 \beta_n e_n \left( \sqrt{g^{-1} f} \right) \, .
\ee
Even in this model, we find Eq.~(\ref{identity1}) again. 
Let us introduce four auxiliary fields $A$, $B$, $C$, $D$ and rewrite the
action into the following form
\be
\label{SactA1b}
S=M_g^2\int d^4x \sqrt{-\det g}\left[ F(A,C) +B(R^{(g)}-A)+ D(T-C) \right]
+2m^2 \int d^4x\sqrt{- \det g}\sum_{n=0}^4 \beta_n e_n
\left( \sqrt{g^{-1} f} \right) \, .
\ee
We again perform the Weyl transformation
(\ref{Weyltr}) in order to transform the action
(\ref{SactA1b}) to the action with  Einstein-Hilbert term.
   Finally we can check the equation of motion with respect to
$A$
\begin{equation}
F_A(A,C)=B \, ,
\end{equation}
and we presume that it can be solved for
$A=\Psi(B,C)$. Putting all these results together we
obtain the action in the form
\begin{align}
S=&M_g^2\int d^4 x \sqrt{-\det g} \left[ R^{(g)} +
\e^{-2\phi}D(\tilde{T}[\e^\phi]-C)
-\frac{2}{3}{g}^{\mu\nu}\nabla_\mu\phi \nabla_\nu\phi-V(\phi,C) \right]
\nonumber \\
&+ 2m^2 \int d^4x \sqrt{-\det g}\sum_{n=0}^4 \beta_n
\e^{\left(\frac{n}{2}-2\right)\phi} e_n
\left( \sqrt{{g}^{-1}f} \right) \, ,
\end{align}
where $\tilde{T}[\e^\phi]$ is the transformed trace of the stress energy tensor
which is model dependent.
Let us now consider the case when we treat $T$ as the external
parameter. Then we can simply solve the equation of motion for $D$
which leads to the replacement $\tilde{T}[\e^\phi]=C$ in the action so that we
have
\be
\label{STextern}
S = M_g^2\int d^4x \sqrt{-\det g} \left[ R^{(g)} -\frac{2}{3}
g^{\mu\nu}\nabla_\mu\phi \nabla_\nu\phi-V(\phi,\tilde{T}[\e^\phi]) \right]
+ 2m^2 \int d^4x \sqrt{-\det g}\sum_{n=0}^4 \beta_n
\e^{\left(\frac{n}{2}-2\right)\phi} e_n
\left(\sqrt{g^{-1}f}\right) \, .
\ee
This action has formally the same form as the action (\ref{act2}) so that
we can quickly say  that given theory is ghost free as well.

The situation will be more complicated in case when $T$ represents
the dynamical quantity. In this case  we should specify its explicit
form in order to perform the Hamiltonian analysis of the coupled
system of massive gravity and the matter that is represented by $T$.
We should also consider the action for the matter field as well. Let
us consider concrete example when the matter is represented by the
scalar field $\psi$ with the action
\begin{equation}
S_\mathrm{matt}=-\int d^4x \sqrt{-\det g} \left[ g^{\mu\nu}\partial_\mu\psi
\partial_\nu\psi+\mF(\psi) \right] \, ,
\end{equation}
where $\mF(\psi)$ is the potential for the scalar field $\psi$. Then
the stress energy tensor has the form
\be
T_{\mu\nu} = -\frac{1}{\sqrt{-\det g}} \frac{\delta S_\mathrm{matt}} {\delta
g^{\mu\nu}} = -\frac{1}{2}g_{\mu\nu} \left[ g^{\rho\sigma}\partial_\rho\psi
\partial_\sigma\psi+\mF(\psi) \right]+\partial_\mu\psi\partial_\nu\psi \, ,
\quad
T=-2\mF-g^{\mu\nu}\partial_\mu\psi\partial_\nu\psi \nonumber \\
\ee
As opposite to the case when $T$ is fixed parameter now we have to
perform the Hamiltonian analysis of the system
\begin{align}
S = & M_g^2\int d^3\bx dt \sqrt{^{(3)}g}N \left[
K_{ij} \mG^{ijkl} K_{kl}+{}^{(3)}R
+ \frac{2}{3}(\nabla_n\phi)^2-\frac{2}{3}g^{ij}\partial_i\phi
\partial_j\phi-V(\phi,C) \right. \nonumber \\
& + \frac{2m^2}{M_g^2} \sum_{n=0}^4 \beta_n \e^{\left(\frac{n}{2}-2\right)\phi}
e_n \left(\sqrt{g^{-1}f} \right) - \frac{1}{M_g^2}
\e^{-2\phi}D(2\mF+C)-\frac{\e^{-2\phi}}{M_g^2}\mF \nonumber \\
&+ \left.\frac{\e^{-\phi}}{M_g^2}(D+1) \nabla_n\psi\nabla_n\psi
-\frac{\e^{-\phi}}{M_g^2}(D+1)g^{ij}\partial_i\psi\partial_j\psi
\right] \, .
\end{align}
This action is the starting point for the Hamiltonian formalism. We
see that in case of massive gravity and the scalar field $\phi$ the
Hamiltonian analysis is the same as in the previous model so that we
will not repeat it here. There are additional terms that arise from
the Hamiltonian analysis of the field $\psi$. We find that there are
additional two primary constraints
\begin{equation}
P_C\approx 0 \, , \quad P_D\approx 0\, ,
\end{equation}
that are variable conjugate to $C$ and $D$, respectively
\begin{equation}
\pb{C(\bx),P_D(\by)}=\delta(\bx-\by) \, , \quad
\pb{D(\bx),P_D(\by)}=\delta(\bx-\by) \, .
\end{equation}
The momentum conjugate to $\psi$ has the form
\begin{equation}
p_\psi=2 \e^{-\phi}\left( D+1 \right)\sqrt{^{(3)}g} \nabla_n\psi  \,
.
\end{equation}
Then it is easy to perform the Legendre transformation with the
resulting Hamiltonian in the form
\be
H = \int d^3\bx \left( N\mC_0+\mH_0+v_Cp_C+v_DP_D \right)
\, ,
\ee
where
\begin{align}
\mC_0 =& \frac{1}{M_g^2\sqrt{^{(3)}g}} \pi^{ij}
\mG_{ijkl}\pi^{kl}-M_g^2\sqrt{^{(3)}g}{}^{(3)}
R^{(g)}+(\mR_k+p_\phi\partial_k\phi+p_\psi\partial_k \psi)\tD^k_{ \
l}\tn^l -2m^2\sqrt{^{(3)}g}\mV \nonumber \\
&+ \frac{3}{8}\frac{1}{\sqrt{^{(3)}g}M_g^2}p_\phi^2
+\frac{2}{3}M_g^2\sqrt{^{(3)}g}g^{ij}
\partial_i\phi\partial_j\phi+M_g^2\sqrt{^{(3)}g}V(\phi,C)
\nonumber \\
&+\frac{\e^{\phi}}{4\sqrt{^{(3)}g}(D+1)}
(p_\psi)^2-\sqrt{^{(3)}g}\e^{-2\phi} (D+1) g^{ij}
\partial_i\psi\partial_j\psi
+ \sqrt{^{(3)}g}\e^{-2\phi}D(2\mF+C)+\sqrt{^{(3)}g}\e^{-2\phi}\mF \, . \nonumber
\\
\mH_0 =& (M\tn^i+L^i) \left(\mR_i+p_\phi\partial_i\phi+p_\psi
\partial_i\psi \right)-2m^2M\sqrt{^{(3)}g} \mU \, .
\end{align}
The preservation of the primary constraints $\pi_N\approx0$,
$\pi_i\approx 0$ again implies an existence of the secondary
constraints $\mC_0$, $\mC_i$ when in $\mC_i$ we have an additional
contribution $\partial_i\psi p_\psi$. On the other hand the
preservation of the constraints $P_C\approx 0$, $P_D\approx 0$
implies following secondary constraints
\be
\mG_C\equiv \frac{\delta \mC_0}{\delta C}\approx 0 \, , \quad
\mG_D\equiv \frac{\delta \mC_0}{\delta D}\approx 0 \, .
\ee
These constraints together with $P_C\approx 0$, $P_D\approx 0$ form
the second class constraints that can be solved for $C$ and $D$ as
functions of canonical variables, at least in principle. Note that
the presence of these constraints does not have consequence for the
existence of the two constraints $\mC_0$, $\mC_0^{II}$ that are
responsible for the elimination of the Boulware-Deser ghost. In
other words, even the non-linear massive  theory where the trace of
the stress-energy tensor is dynamical variable is ghost-free.

\section{Discussion}
In summary, using very clever ghost-free construction for massive
generalization of General Relativity we propose the elegant way to
formulate the class of ghost-free massive $F(R)$ gravities. The
hamiltonian formulation of this theory is developed using its presentation
in scalar-tensor form as well as analogy with the hamiltonian treatment of
usual ghost-free massive gravity. Based on this analogy we prove that our
theory turns out to be also ghost-free. The same strategy is applied for
generalization of $F(R,T)$ gravity. Again, we demonstrate that its massive
version turns out to be ghost-free.

Furthermore, the cosmological evolution in massive $F(R)$ gravity under
consideration is studied. It turns out that Bianchi identity gives an
equation which constrains the cosmic evolution quite strongly if compare
with the case of convenient $F(R)$ theory where no such constraint
appears.
Nevertheless, the possibility of cosmic acceleration (especially, in the
presence of cold dark matter) is established. Of course, the ocurrence of
cosmic acceleration in such theory is much restricted if compare even with
  massive $F(R)$ bigravity.
Nevertheless, adding extra scalar fields in analogy with massive $F(R)$
bigravity may improve the ocurrence of accelerated expansions.
  Using proposed
strategy one can generate more massive extensions of modified gravity
theories. This will be discussed elsewhere.

\section*{Acknowledgments}

This work is supported in part by  MINECO (Spain), FIS2010-15640, AGAUR
(Generalitat de Catalunya),
contract 2009SGR-345, and MES project 2.1839.2011 (Russia) (S.D.O.) and
  by the JSPS Grant-in-Aid for Scientific Research (S) \# 22224003
and (C) \# 23540296 (S.N.). The work of J.K.
 is supported by the Grant agency of the Czech republic under the grant
P201/12/G028.

\appendix

\section{Curvature in FRW metric with conformal time}

We now give the explicit forms of connections and curvature in the FRW metric
with conformal time
in (\ref{Fbi10}).

The non-vanishing components of the connections $\Gamma^\mu_{\nu\rho}$ are
given by
\be
\label{A1}
\Gamma^t_{tt} = H\, ,\quad \Gamma^t_{ij} = H \delta_{ij}\, ,\quad
\Gamma^i_{tj} = \Gamma^i_{jt} = H \delta^i_{\ j}\, .
\ee
The non-vanishing components of the Ricci curvature and the scalar curvatures
are given by
\be
\label{A2}
R_{yy} = - 3\dot H\, ,\quad
R_{ij} = \left( \dot H + 2 H^2 \right) \delta_{ij}\, ,\quad
R = \frac{6}{a^2} \left( \dot H + H^2 \right)\, .
\ee


\begin{thebibliography}{99}


\bibitem{Fierz:1939ix}
M.~Fierz and W.~Pauli,
Proc.\ Roy.\ Soc.\ Lond.\ A {\bf 173} (1939) 211.

\bibitem{Hinterbichler:2011tt}
K.~Hinterbichler,
Rev.\ Mod.\ Phys.\  {\bf 84} (2012) 671
[arXiv:1105.3735 [hep-th]].

\bibitem{Boulware:1974sr}
D.~G.~Boulware and S.~Deser,
Annals Phys.\  {\bf 89} (1975) 193.


\bibitem{Boulware:1973my}
D.~G.~Boulware and S.~Deser,
Phys.\ Rev.\ D {\bf 6} (1972) 3368.

\bibitem{vanDam:1970vg}
H.~van Dam and M.~J.~G.~Veltman,
Nucl.\ Phys.\ B {\bf 22} (1970) 397; \\
V.~I.~Zakharov,
JETP Lett.\  {\bf 12} (1970) 312
[Pisma Zh.\ Eksp.\ Teor.\ Fiz.\  {\bf 12} (1970) 447].

\bibitem{Vainshtein:1972sx}
A.~I.~Vainshtein,
Phys.\ Lett.\ B {\bf 39} (1972) 393.

\bibitem{Luty:2003vm}
M.~A.~Luty, M.~Porrati and R.~Rattazzi,
JHEP {\bf 0309}, 029 (2003)
[hep-th/0303116]; \\
A.~Nicolis and R.~Rattazzi,
JHEP {\bf 0406} (2004) 059  [hep-th/0404159].

\bibitem{Dvali:2000hr}
G.~R.~Dvali, G.~Gabadadze and M.~Porrati,
Phys.\ Lett.\ B {\bf 485} (2000) 208
[hep-th/0005016]; \\
C.~Deffayet,
Phys.\ Lett.\  B {\bf 502}, 199 (2001)
[arXiv:hep-th/0010186]; \\
C.~Deffayet, G.~R.~Dvali and G.~Gabadadze,
Phys.\ Rev.\ D {\bf 65}, 044023 (2002)
[astro-ph/0105068].

\bibitem{deRham:2010ik}
C.~de Rham and G.~Gabadadze,
Phys.\ Rev.\ D {\bf 82}, 044020 (2010)
[arXiv:1007.0443 [hep-th]]. 

\bibitem{deRham:2010kj}
C.~de Rham, G.~Gabadadze and A.~J.~Tolley,
Phys.\ Rev.\ Lett.\  {\bf 106} (2011) 231101
[arXiv:1011.1232 [hep-th]].

\bibitem{Hassan:2011hr}
S.~F.~Hassan and R.~A.~Rosen,
Phys.\ Rev.\ Lett.\  {\bf 108} (2012) 041101
[arXiv:1106.3344 [hep-th]].

\bibitem{Hassan:2011zd}
S.~F.~Hassan and R.~A.~Rosen,
JHEP {\bf 1202} (2012) 126
[arXiv:1109.3515 [hep-th]].

\bibitem{Hassan:2011tf}
S.~F.~Hassan, R.~A.~Rosen and A.~Schmidt-May,
JHEP {\bf 1202} (2012) 026
[arXiv:1109.3230 [hep-th]].

\bibitem{Hassan:2011vm}
S.~F.~Hassan and R.~A.~Rosen,
JHEP {\bf 1107} (2011) 009
[arXiv:1103.6055 [hep-th]].

\bibitem{Kluson:2012zz} 
J.~Kluson,
arXiv:1209.3612 [hep-th].
S.~F.~Hassan, A.~Schmidt-May and M.~von Strauss,
arXiv:1203.5283 [hep-th]; \\
K.~Koyama, G.~Niz and G.~Tasinato,
Phys.\ Rev.\ D {\bf 84} (2011) 064033
[arXiv:1104.2143 [hep-th]]; arXiv:1210.4378;\\
G.~D'Amico, C.~de Rham, S.~Dubovsky, G.~Gabadadze, D.~Pirtskhalava and
A.~J.~Tolley,
Phys.\ Rev.\ D {\bf 84} (2011) 124046
[arXiv:1108.5231 [hep-th]]; \\
K.~Hinterbichler and R.~A.~Rosen,
JHEP {\bf 1207} (2012) 047
[arXiv:1203.5783 [hep-th]]; \\
V.~Baccetti, P.~Martin-Moruno and M.~Visser,
arXiv:1205.2158 [gr-qc]; \\
T.~Kobayashi, M.~Siino, M.~Yamaguchi and D.~Yoshida,
arXiv:1205.4938 [hep-th]; \\
K.~Nomura and J.~Soda,
Phys.\ Rev.\ D {\bf 86} (2012) 084052
[arXiv:1207.3637 [hep-th]]; \\
E.~N.~Saridakis,
arXiv:1207.1800 [gr-qc]; \\
Y.~-F.~Cai, C.~Gao and E.~N.~Saridakis,
JCAP {\bf 1210} (2012) 048
[arXiv:1207.3786 [astro-ph.CO]]; \\
Y.~-l.~Zhang, R.~Saito and M.~Sasaki,
JCAP {\bf 1302} (2013) 029
[arXiv:1210.6224 [hep-th]]; \\
M.~Mohseni,
JCAP {\bf 1211} (2012) 023
[arXiv:1211.3501 [hep-th]]; \\
K.~Hinterbichler, J.~Stokes and M.~Trodden,
Phys.\  Lett.\ B {\bf 725} (2013) 1
[arXiv:1301.4993 [astro-ph.CO]]; \\
M.~Andrews, G.~Goon, K.~Hinterbichler, J.~Stokes and M.~Trodden,
Phys.\ Rev.\ Lett.\  {\bf 111} (2013) 061107
[arXiv:1303.1177 [hep-th]]; \\
R.~Gannouji, M.~W.~Hossain, M.~Sami and E.~N.~Saridakis,
Phys.\ Rev.\ D {\bf 87} (2013) 123536
[arXiv:1304.5095 [gr-qc]]; \\
S.~Capozziello and P.~Martin-Moruno,
Phys.\ Lett.\ B {\bf 719} (2013) 14
[arXiv:1211.0214 [gr-qc]]; \\
G.~Leon, J.~Saavedra and E.~N.~Saridakis,
Class.\ Quant.\ Grav.\  {\bf 30} (2013) 135001
[arXiv:1301.7419 [astro-ph.CO]].

\bibitem{Kluson:2012wf}
J.~Kluson,
Phys.\ Rev.\ D {\bf 86} (2012) 044024
[arXiv:1204.2957 [hep-th]].

\bibitem{Hassan:2011ea}
S.~F.~Hassan and R.~A.~Rosen,
JHEP {\bf 1204} (2012) 123
[arXiv:1111.2070 [hep-th]].

\bibitem{Damour:2002wu}
T.~Damour, I.~I.~Kogan and A.~Papazoglou,
Phys.\ Rev.\ D {\bf 66} (2002) 104025
[hep-th/0206044].

\bibitem{Volkov:2011an}
M.~S.~Volkov,
JHEP {\bf 1201} (2012) 035
[arXiv:1110.6153 [hep-th]]. 

\bibitem{vonStrauss:2011mq}
M.~von Strauss, A.~Schmidt-May, J.~Enander, E.~Mortsell and S.~F.~Hassan,
JCAP {\bf 1203} (2012) 042
[arXiv:1111.1655 [gr-qc]].

\bibitem{Berg:2012kn}
M.~Berg, I.~Buchberger, J.~Enander, E.~Mortsell and S.~Sjors,
JCAP {\bf 1212} (2012) 021
[arXiv:1206.3496 [gr-qc]].

\bibitem{Nojiri:2012zu}
S.~Nojiri and S.~D.~Odintsov,
Phys.\ Lett.\ B {\bf 716} (2012) 377,
arXiv:1207.5106 [hep-th].

\bibitem{Nojiri:2012re}
S.~Nojiri, S.~D.~Odintsov and N.~Shirai,
JCAP {\bf 1305} (2013) 020
[arXiv:1212.2079 [hep-th]].

\bibitem{Cai:2013lqa}
Y.~-F.~Cai, F.~Duplessis and E.~N.~Saridakis,
arXiv:1307.7150 [hep-th].

\bibitem{Hassan:2012qv}
S.~F.~Hassan, A.~Schmidt-May and M.~von Strauss,
Phys.\ Lett.\ B {\bf 715} (2012) 335
[arXiv:1203.5283 [hep-th]].

\bibitem{Golovnev2:2011aa}
A.~Golovnev,
Phys.\ Lett.\ B {\bf 707} (2012) 404
[arXiv:1112.2134 [gr-qc]].

\bibitem{Huang:2013mha}
Q.~-G.~Huang, K.~-C.~Zhang and S.~-Y.~Zhou,
arXiv:1306.4740 [hep-th].

\bibitem{Harko:2011kv}
T.~Harko, F.~S.~N.~Lobo, S.~Nojiri and S.~D.~Odintsov,
Phys.\ Rev.\ D {\bf 84} (2011) 024020
[arXiv:1104.2669 [gr-qc]].

\bibitem{Nojiri:2010wj}
S.~Nojiri and S.~D.~Odintsov,
Phys.\ Rept.\  {\bf 505} (2011) 59
[arXiv:1011.0544 [gr-qc]]; 
eConf C {\bf 0602061} (2006) 06
[Int.\ J.\ Geom.\ Meth.\ Mod.\ Phys.\  {\bf 4} (2007) 115]
[hep-th/0601213].

\bibitem{Capozziello:2010zz}
S.~Capozziello and V.~Faraoni,
``Beyond Einstein gravity: A Survey of gravitational theories
for cosmology and astrophysics,'', Springer 2010,
DOI: 10.1007/978-94-007-0165-6; \\
S.~Capozziello and M.~De Laurentis,
Phys.\ Rept.\  {\bf 509} (2011) 167
[arXiv:1108.6266 [gr-qc]].

\bibitem{Gourgoulhon:2007ue}
E.~Gourgoulhon,
gr-qc/0703035.

\bibitem{Arnowitt:1962hi}
R.~L.~Arnowitt, S.~Deser, C.~W.~Misner,
[gr-qc/0405109].

\end{thebibliography}
\end{document}